\documentstyle[prl,aps,multicol,graphicx]{revtex}

\newcommand{\be}{\begin{equation}}
\newcommand{\ee}{\end{equation}}

\newcommand{\bfq}{{\bf q}}
\newcommand{\bfp}{{\bf p}}

\newcommand{\pise}{$\Pi_{se}\;$}
\newcommand{\piex}{$\Pi_{ex}\;$}
\newcommand{\pifl}{$\Pi_{fl}\;$}
\newcommand{\pis}{$\Pi^{*}\;$}
\newcommand{\piz}{$\Pi^{0}\;$}
\newcommand{\gam}{$\Gamma(q)\;$}
\newcommand{\kt}{\tilde{k}}
\newcommand{\qt}{\tilde{q}}
\newcommand{\pt}{\tilde{p}}
\newcommand{\three}{$\Lambda^{(3)}(\qt,\pt) \; \;$}
\newcommand{\bleq}{\ifpreprintsty
                   \else
                   \end{multicols}\vspace*{-3.5ex}{\tiny
                   \noindent\begin{tabular}[t]{c|}
                   \parbox{0.493\hsize}{~} \\ \hline \end{tabular}}
                   \fi}
\newcommand{\eleq}{\ifpreprintsty
                   \else
                   {\tiny\hspace*{\fill}\begin{tabular}[t]{|c}\hline
                    \parbox{0.49\hsize}{~} \\
                    \end{tabular}}\vspace*{-2.5ex}\begin{multicols}{2}
                    \fi}
\newcommand{\bcols}{\ifpreprintsty\else\begin{multicols}{2}\fi}
\newcommand{\ecols}{\ifpreprintsty\else\end{multicols}\fi}
\newcommand{\lam}{$\Lambda^{(3)}$}
\newlength{\parindlen}
\newcommand{\parin}{\hspace{\parindent}}

\begin{document}

%\twocolumn[\hsize\textwidth\columnwidth\hsize\csname @twocolumnfalse\endcsname

\title{Singular Structure and Enhanced Friedel Oscillations in the 
Two-Dimensional Electron Gas}

\author{I. G. Khalil*, N.~W.~Ashcroft$\dagger$, and M. Teter$\dagger$}

\address{*Gene Network Sciences, 2359 N Triphammer Rd, Ithaca, NY 14850\\
$\dagger$Cornell Center for Materials Research, and 
the Laboratory of Atomic and Solid State Physics, 
Cornell University,  Ithaca, NY 14853-2501}

\date{\today}

\maketitle

\bigskip

\begin{abstract}

\noindent We calculate the leading order corrections (in $r_s$) to the
static polarization $\Pi^{*}(q,0,)$, with dynamically 
screened interactions, for the two-dimensional electron gas. 
The corresponding diagrams all exhibit singular logarithmic behavior in 
their derivatives
at $q=2 k_F$ and provide significant enhancement to the proper
polarization particularly 
at low densities.  At a density of $r_s=3$, the contribution from the leading order
{\em fluctuational} diagrams exceeds both the zeroth order (Lindhard) response
and
the self-energy and exchange contributions.  We comment on the
importance of these diagrams  
in two-dimensions and make comparisons to an equivalent 
three-dimensional electron gas;  we also consider the impact these finding 
have on $\Pi^{*}(q,0)$ computed to all orders in perturbation theory.
\end{abstract}
\vskip 0.5in

%\narrowtext 
%]

\bcols

\begin{figure}
\centerline{\includegraphics[width=8.5cm]{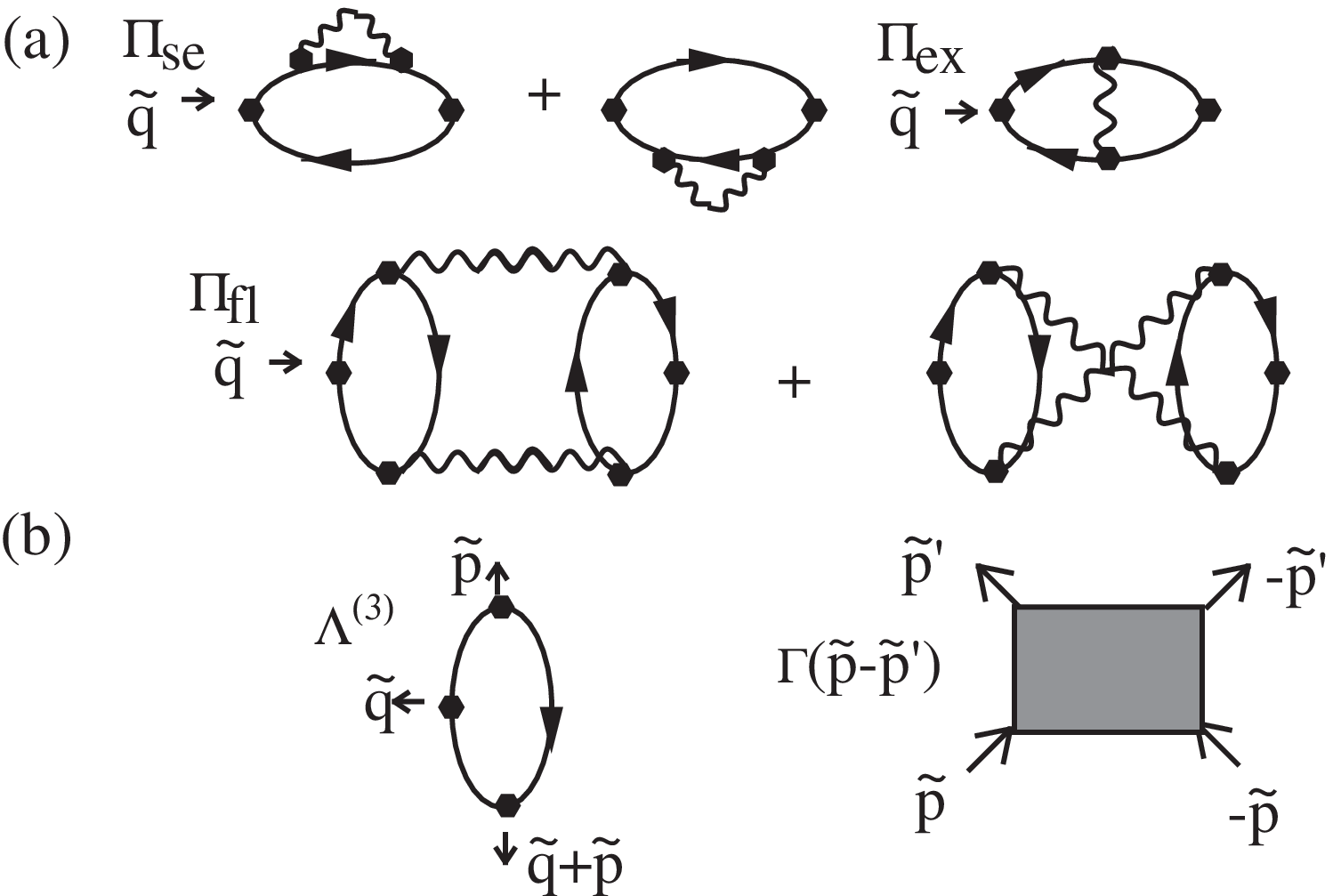}}
\refstepcounter{figure}
{\noindent \small FIG.~\ref{figure1}. 
(a) Leading order corrections to the proper polarization $\Pi^{*}$.
  The interaction lines represent RPA screened Coulomb interactions.
(b) The three-point diagram \lam and the effective scattering vertex \gam.
\protect}
\label{figure1}
\end{figure}

\parin Two-dimensional electron systems with standard singular Coulombic
interactions have attracted a great deal of attention in recent years.
Of particular interest have been the driving mechanisms behind 
the high temperature superconductors, the metal-insulator transition
exhibited by the two-dimensional localized electrons in semiconductors, 
and the novel charge and spin-density-wave ordering found in layered compounds. 
%(XXXsite references)
An important question is the degree to which we might attribute this 
novel behavior specifically to the two-dimensionality and the 
singular character of the Coulomb interaction?  
In this letter, we attempt to provide an answer 
by computing the leading order corrections 
to the proper static polarization $\Pi^{*}$ beyond the 
random-phase approximation 
(RPA) through inclusion of the diagrams 
presented in Fig.~\ref{figure1} shown there as the self-energy, 
exchange, and
%%%%%%%%%%%%%%%%%%%%%%%%%%%%%%%%% 
%%%%%%%%%%%%%%%%%%%%%%%%%%%%%%%%%%%%%%%
\noindent fluctuation diagrams, respectively.  For the
three-dimensional electron 
gas, these diagrams have 
been well studied, and in particular have been computed and analyzed in detail
(see \cite{langreth}-\cite{ra} and references therein).  For a strictly
two-dimensional electron gas with $2\pi e^2/q$ interaction, we report here
that the response function beyond the RPA exhibits singular 
structure that is far more pronounced than found from the
zeroth order Lindhard response function; these leading order corrections
play an important role even at high-densities.  

\parin It is not surprising that
all three leading order diagrams exhibit similar singular structure since
they can be reduced to expressions containing the three-point 
diagram shown \lam in Fig.~\ref{figure1} (multiplied by Coulombic interaction 
lines).  This singular structure should extend to all orders in perturbation theory 
since all higher order corrections to the polarization also reduce to a sum of three-point 
diagrams \lam (again multiplied by Coulombic interaction lines).
More importantly, the enhancement of the polarization from all of three of the 
leading order diagrams is already significant
at high densities, but as the density is reduced the contribution
from the fluctuation diagrams rapidly exceeds that of the self-energy and exchange
contributions.  This is in marked contrast to the three-dimensional electron gas 
where fluctuational diagrams provide a small enhancement to the 
corresponding zeroth order Lindhard response when compared to contributions
solely from the exchange and self-energy diagrams 
(for densities in the metallic range\cite{ra}).  
The increased role that fluctuations are expected to play in two 
dimensions is evident.  

\parin Begin with the static 
zeroth-order Lindhard response function $\Pi^{0}$, appropriate to
the random-phase approximation.  
The well-known singular behavior of  $\Pi^{0}$ leads to possible instabilities 
arising from electron-electron interactions.
In three dimensions, this singularity also leads to a modulation 
in the long range effective electron-electron interaction which in
three dimensions takes the form 
$V(r)\sim cos(2 k_F r)/r^3$. Based on this Kohn and 
Luttinger \cite{kohn} were the first to consider the possibility of 
\noindent ground-state 
electron pairing in the presence of 
purely repulsive forces where they found attractions in the effective
vertex $\Gamma(q)$ of Fig.~\ref{figure1} for large 
angular momentum quantum numbers.  Details of 
the shape of the Fermi surface (for example, nesting)
might well enhance such possibilities.

\parin In two dimensions, and at the level of the Lindhard 
response function, these effects are not especially 
prominent.  The function
$\Pi^{0}(q)$ is then a constant for $q \leq 2k_F$, 
and exhibits a square root singularity in the derivative
for momentum $q >2 k_F$\cite{stern}.  
Because of this arguments for
possible charge/spin instabilities based on $\Pi^{0}$ are not
compelling.  While this square root singularity does lead
to a long range component in the effective interaction 
($V(r)\sim sin(2 k_F r)/r^2$) for two dimensions, the effective
interaction vertex $\Gamma(q)$ of Fig.~\ref{figure1} computed at this low order
remains independent of momentum at the Fermi surface.  Only an s-wave
repulsive scattering channel opens at this order and higher-order
effects must be considered in order to find attractive scattering
channels\cite{chubukov}.  But beyond the random phase approximation (RPA) the 
situation is quite different and we
report here that the first order corrections
to $\Pi^{*}$ already give singularities (in part logarithmic) in the derivative on 
both sides of $2 k_F$, and this singular structure is further enhanced according to
the number of three-point diagrams \lam and interaction lines 
appearing in a polarization diagram.  
As an example, the fluctuation diagrams \pifl
of Fig.~\ref{figure1} contains two three-point diagrams and two dynamically 
screened interaction lines, and shows greater 
singular behavior at the Fermi surface in two-dimensions 
than the self-energy and exchange diagrams 
which depend on a single three-point diagram and single dynamically screened interaction
line (see Eqs.~\ref{eq:pise} and~\ref{eq:piex}).  
Consequently the enhancement near $2 k_F$
in $\Pi^{*}$ from these diagrams leads
to a greater attraction  in {\em real space} for the screened interaction 
$V_q^{*}=v_q/(1-v_q\Pi^{*})$  when compared with the RPA effective interaction 
$V_q^{RPA}=v_q/(1-v_q\Pi^{0})$.

\parin As noted in references 
\cite{ra} and \cite{cenni}, the self energy \pise, exchange \piex, 
and fluctuation \pifl diagrams depicted in Fig.~\ref{figure1} can all be written 
in terms of the three-point function; thus
\bleq
\begin{eqnarray}
\Pi_{se}(\pt)&=& Tr_{\qt} v_{RPA}(\qt) \times \left[ 
\frac{\partial}{\partial(i\omega_q)}-\frac{\partial}{\partial(i\omega_p)} 
\right]
\times 
\left[ \Lambda^{(3)}(\pt,\qt)+\Lambda^{(3)}(\qt,\pt) \right], 
\label{eq:pise}
\\
\Pi_{ex}(\pt)&=& Tr_{\qt} \frac{v_{RPA}(\qt)}{\bfp \cdot \bfq/m} 
\times 
\left[ \Lambda^{(3)}(\pt,\qt)+\Lambda^{(3)}(\qt,\pt)-\Lambda^{(3)}(-\pt,\qt)
-\Lambda^{(3)}(\qt,-\pt) \right],
\label{eq:piex}
\\
\Pi_{fl}(\pt)&=&-\frac{1}{2} Tr_{\qt} v_{RPA}(\qt) v_{RPA}(\pt-\qt)
 \times 
\left[ \Lambda^{(3)}(\qt,\pt-\qt)+\Lambda^{(3)}(\pt-\qt,\qt) \right] ^2.
\label{eq:pifl}
\end{eqnarray}
\eleq
\noindent Here $\pt=(i\omega_p,\bfp)$ is the energy-momentum variable,  
$v_{RPA}$ is the screened Coulombic interaction 
in the random phase approximation (using
$v_q=2\pi e^2/q$), and $Tr _{\qt}$ stands for the two-dimensional
trace over all momentum and energy variables, i.e. 
$\int d\omega_q/(2\pi) \int d^2q/(2\pi)^2$.
The requisite three-point function \three, is given by
\begin{eqnarray}
\Lambda^{(3)}(\qt,\pt)=-2 Tr_{\kt} G_0(\kt) G_0(\kt+\qt) G_0(\kt+\qt+\pt),
\label{eq:lam}
\end{eqnarray}
where exact analytical expressions for two-dimensional forms 
are given by Neumayr and Metzner\cite{neumayr}.
We compute the diagrams of Fig.~\ref{figure1} in imaginary frequency space to circumvent nonintegrable 
divergences near $2 k_F$ and also when integrating through plasmon peaks in 
the screened interaction $v_{RPA}(\qt)$.  
To obtain the static polarizability we simply perform the requisite
analytic continuation $\Pi(p,i\omega_p) \rightarrow \Pi(p,\omega+i\delta)=
\Pi(p,0+i\delta)$.
 
\begin{figure}
\centerline{\includegraphics[width=8.5cm]{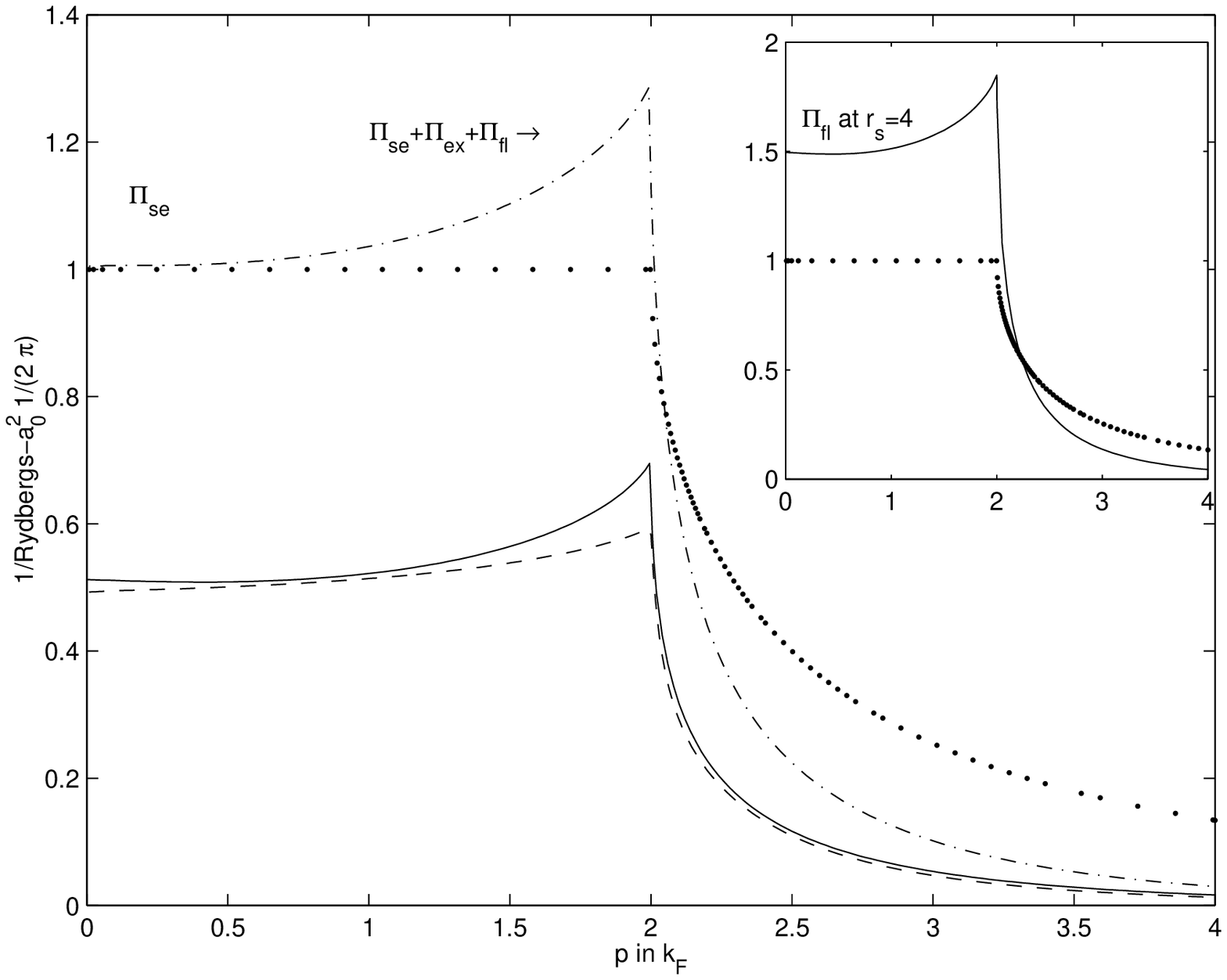}}
\refstepcounter{figure}
%\smallskip
{\noindent \small FIG.~\ref{pstar}.
The solid curve is the static \pifl, the dashed curve shows $(\Pi_{se}+\Pi_{ex})$,
 and the dashed-dotted curve is the sum of all three of the first order correction all 
at $r_s=2$.  The inset contains the fluctuation diagram at a lower density of 
$r_s=4$.  The dotted curve is the 2D Lindhard response.
\protect
}
\label{pstar}
\end{figure}

\parin The numerical procedure for computing \pise, \piex, and \pifl is 
lengthy but straightforward.  The results for a density of $r_s=2$ (not a 
particularly high density) are plotted 
in Fig.~\ref{pstar} and can be 
summarized as follows:  First,
\piex exceeds \piz for momenta up to and slightly above $2 k_F$, but 
\pise and \piex have opposite signs and their sum largely cancels.  
However, \pifl at this density, already 
gives a slightly larger contribution than \pise and \piex combined.  
Second, all three diagrams are finite but discontinuous
in slope at $2 k_F$; their derivatives diverge logarithmically 
at all densities.  We have determined through a numerical fit that the logarithmic 
structure is of the form $\sqrt{|q-2 k_F|} \log|q- 2 k_F|$ on either side of
$q=2 k_F$.  Maldague \cite{maldague} 
has numerically computed the 2D self-energy and exchange contributions 
in the absence of screening 
and found similar logarithmic divergence in the slope for $q > 2 k_F$ 
by invoking
an electrostatic analogy that treats the integrals over
$(\Pi_{se}+\Pi_{ex})$ as equivalent charge distributions of radius $k_F$ concentrated near the origin
times the 2D bare interaction.  The divergence in the slope is stronger
above $2 k_F$ in overall magnitude than below $2 k_F$ due to the
analytic form of \lam which contains non-trivial expressions involving square root 
singularities (similar to that found in \piz).  These expressions contribute to the analytic form
of \lam above, but not below $2 k_F$.
Overall the divergent structure is greatest for \pifl as stated earlier.  
The contribution from all three diagrams is significant already at $r_s=2$.  The role 
of fluctuations  
%%%%%%%%%%%%%%%%%%%%%%%%%%%%%%%%%%%%%%%%%%%%%%%%%%%%%%%%%%%%%%%%%
%%%%%%%%%%%%%%%%%%%%%%%%%%%%%%%%%%%%%%%%%%%%%%%%%%%%%%%%%%%%%%%%%%%%%%55
\noindent is expected to progressively
dominate with decreasing density, and this is apparent at
$r_s=3$ where the fluctuation diagrams begin 
to exceed \piz.  By $r_s=8$ (a typical density for high-temperature
superconductors or the two-dimensional electron(hole) gases in MOSFETS)
it is a factor of four greater than \piz where at this density the sum of the
self-energy and exchange diagrams only just begins to exceed the zeroth order 
Lindhard response.  Overall the effect of dynamical screening as $r_s$ increases
is to reduce the total enhancement of the polarization diagrams as well as the degree
of divergence in the slope (see \cite{maldague} for a comparison with the self-energy and 
exchange diagrams computed with the bare interaction).

\begin{figure}
\centerline{\includegraphics[width=8.5cm]{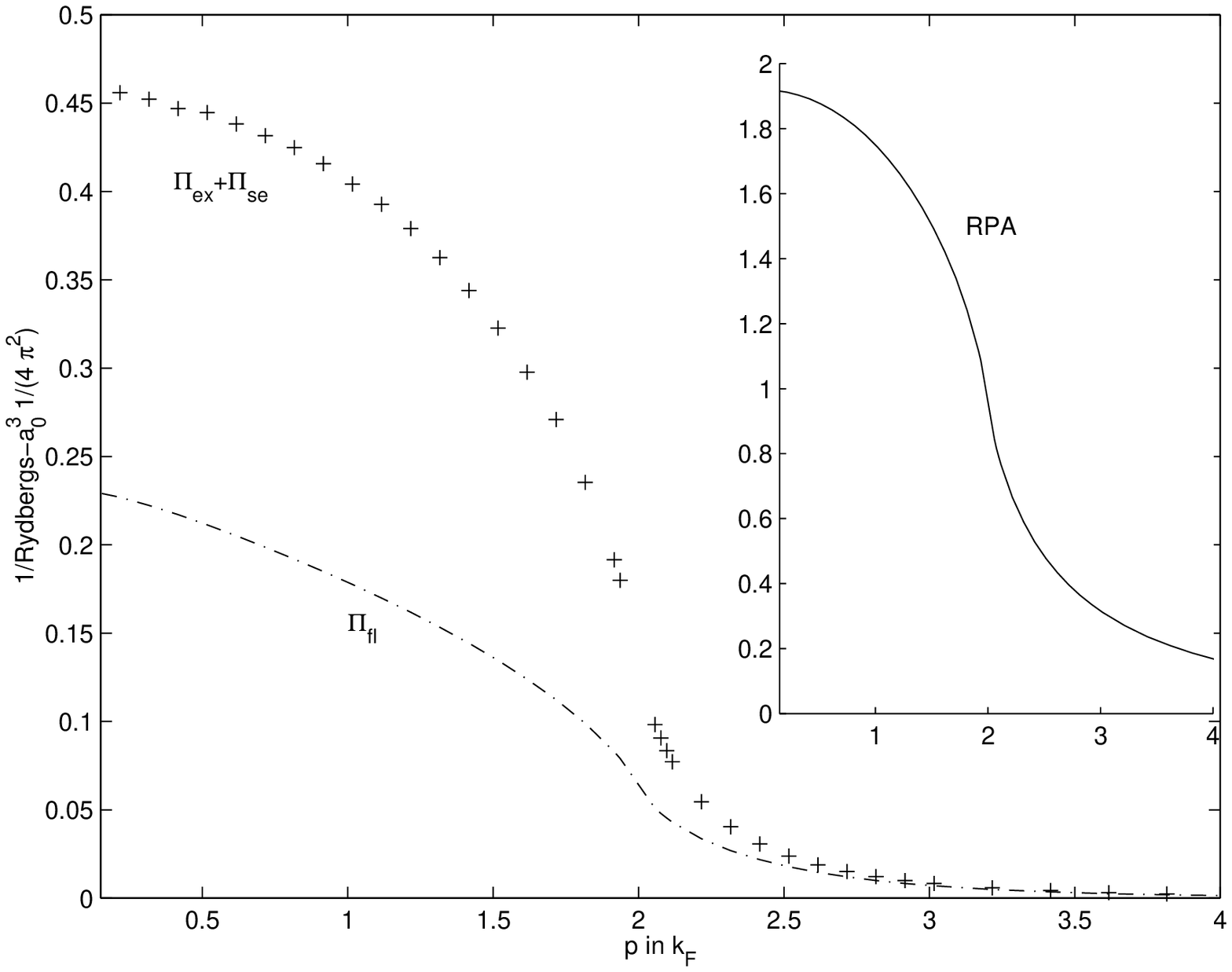}}
\refstepcounter{figure}
%\smallskip
%\vspace{-2.4cm}
{\noindent \small FIG.~\ref{3d}.
The Static -\pise+\piex, and \pifl in 3D at $r_s=2$. The inset shows 
the 3D Lindhard response.\protect}
\label{3d}
\end{figure}

\parin The leading order corrections to the polarization beyond the RPA 
are not as significant in three-dimensions.  
In the absence of screening the  enhancement
$(\Pi_{se}+\Pi_{ex})/\Pi^{0}$ is $0.17 r_s$ at $p=0$ compared with $0.45 r_s$ in 2D
, and then proceeds to sharply falls off
to half its value at $2 k_F$ \cite{engel}.  The singular structure found in 
these diagrams resembles that of the 3D Lindhard response, but with dynamical
screening the overall enhancement of these diagrams is greatly reduced as shown
in Fig.~\ref{3d} (the data is from reference \cite{ra}).  This data is representative
of the entire metallic density range and confirms the expectation
that self-energy and exchange effects are more pronounced in 2D.  
More striking is the significant enhancement of \piz in 2D from the 
fluctuation diagrams
as seen by comparing the plots of Fig.~\ref{pstar} with the
three-dimensional data of Fig.~\ref{3d}.  Physically, the diagrams of \pifl 
can be thought of as representing quantum  fluctuations in 
the screening clouds 
surrounding two interacting 
electrons, and we expect such effects to play a greater 
%%%%%%%%%%%%%%%%%%%%%%
%%%%%%%%%%%%%%%%%%%%%%%
\noindent role in 2D\cite{rapcewicz,cornu}.  
Enhanced fluctuations in two-dimensions are responsible for the destruction of
superconducting phase coherence and long range crystalline order for $T>0$, 
and thus it is not surprising to see this theme played out here as well
\cite{hohenberg,mermin}.

\parin As mentioned earlier, the general structure of these leading
order diagrams is an integral over three-point functions \lam
augmented by Coulombic interaction lines.  For higher orders, 
an explicit reduction formula
derived by Neumayr and Metzner\cite{neumayr} holds that the general N-loop diagram can
be expressed in terms of 3-loop diagrams over an appropriate
energy/momentum transfer factor  (i.e. in \piex that factor is $(\bfp \cdot \bfq)$
in Eq.(\ref{eq:piex})) and thus the structure of all
higher order diagrams will be of similar form (for an explicit formula
see reference\cite{neumayr}).  We therefore expect the singular 
logarithmic features
found in these lower order diagrams to be present in {\em all} higher order corrections.  Cancellations between some of the higher order diagrams 
may well reduce the overall singular behavior of the total proper 
polarization $\Pi^{*}$.
  
\parin As further higher order corrections are considered there will continue to be 
significant cancellation between vertex and self-energy diagrams in 
such a way that their sum makes a smaller net contribution, and any reasonable 
higher order calculation to \pis should include the right mix of vertex and self-energy 
diagrams(as is the case for the conserving approximation of
Baym and Kadanoff, for example)\cite{mahan}.  But the contribution from diagrams
of the fluctuation type differs in two respects.  First, the internal Coulomb lines 
are screened and already represent the sum of an entire class of diagrams in a 
Dyson series sense\cite{rapcewicz}; second, higher order contributions
of diagrams from the fluctuational type do not cancel but actually sum
to contribute to quantities such as the pair correlation function at zero 
separation, $g(0)$\cite{gold}.  We generally expect, therefore, that contributions 
from all 
%%%%%%%%%%%%%%%%%%%%%%%%%%%%%%%
\begin{figure}
\centerline{\includegraphics[width=8.5cm]{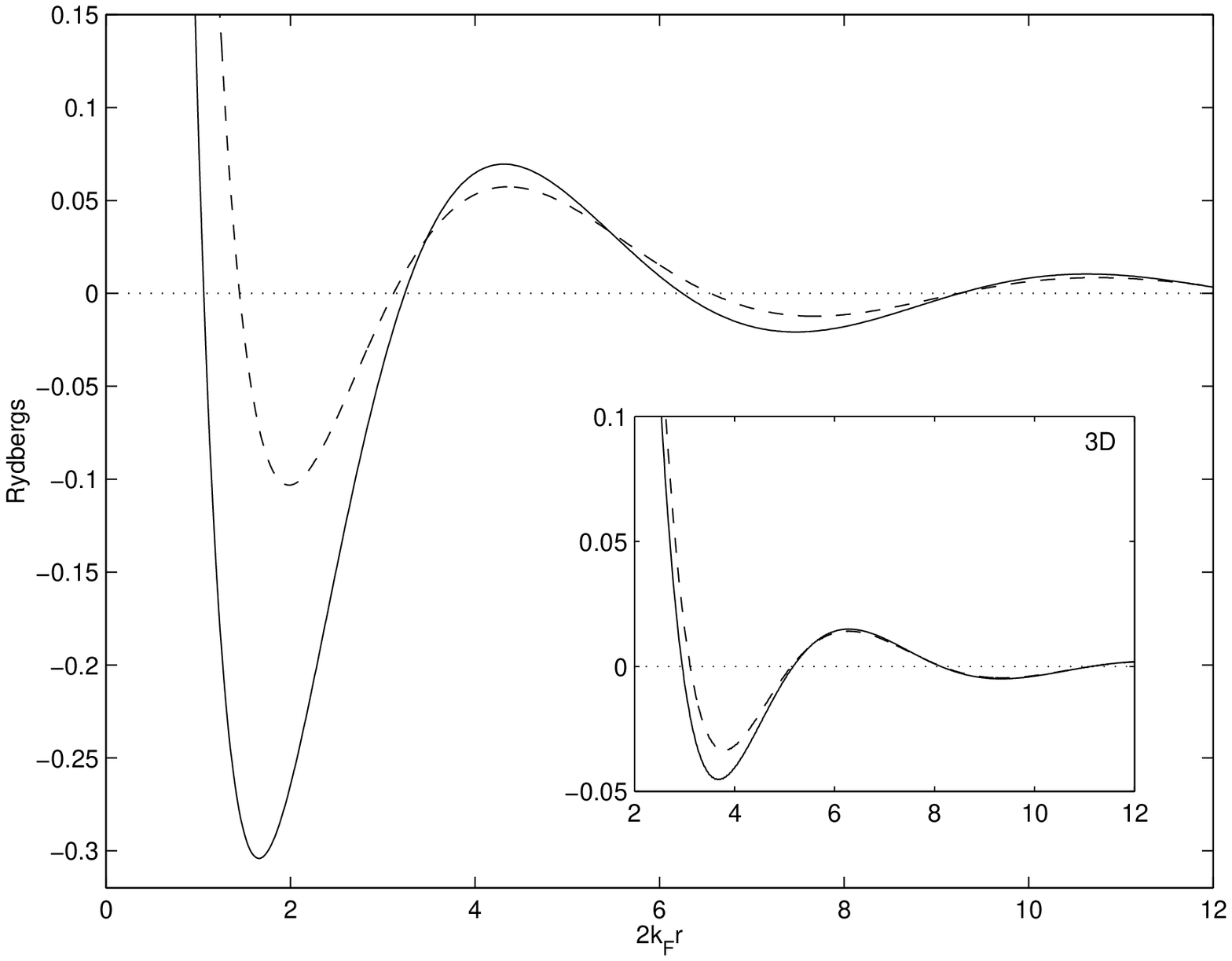}}
\refstepcounter{figure}
%\smallskip
{\noindent \small FIG.~\ref{eff}.
The screened effective interaction in real space showing
an enhancement in the amplitude of the Friedel Oscillations due to the singular structure
in \pis at $r_s=2$ (the dashed line is the RPA screened 
interaction).  The inset shows the three dimensional case at 
an equivalent density.\protect}
\label{eff}
\end{figure}
%%%%%%%%%%%%%%%%%%%%%%%%5
\noindent such diagrams (e.g. the ladder series) will significantly 
contribute 
to the total proper polarization $\Pi^{*}$ leading to an overall 
enhancement and to
singular logarithmic 
\noindent structure at $2 k_F$.

\parin Finally, we return to the important Kohn-Luttinger question for the 
two dimensional case, but in the context of the leading
order corrections of Fig.~\ref{figure1}.  The effective interaction
vertex $\Gamma(q)$ in the case of the electron gas must 
include a sum of all polarization diagrams due to the singularity at small 
momentum transfer $\bfq=\bfp-\bfp'$.  If we include the leading
order contributions to $\Gamma(q)$ we obtain
$\Gamma(q)=v_q/(1-v_q \Pi^*(q))$
with $\Pi^{*}=\Pi^{0}+\Pi_{se}+\Pi_{ex}+\Pi_{fl}$.   
The logarithmic singularity in $\Pi^{*}$ now gives rise to a an attraction 
in $\Gamma_l$ for even angular momentum $l$ starting at $l=2$ 
(${\em d}-wave$), behavior
not possible if only \piz entered into the argument.  
These singularities also significantly 
contribute to the modulation in real space of the equivalent effective interaction 
(i.e. $V(q)=v_q/(1-v_q \Pi^{*})$)
now plotted in Fig.~\ref{eff} for $r_s=2$.  
The principal depth (the first minimum) of the effective interaction
is striking, and the effect is further enhanced as
the fluctuation diagrams exceed $\Pi^{0}$ for densities as low as $r_s=3$.
The inset in Fig.~\ref{eff} shows the three-dimensional case for comparison 
where the singularity at $2 k_F$ from the leading order corrections leads to 
only a small enhancement of the Friedel oscillations.  

\parin To summarize, the low order diagrams, reducible
to similar integral forms, all exhibit
singular logarithmic behavior at $2 k_F$ and provide significant enhancement
over the zeroth order Lindhard response.  
Real systems either have either a finite transverse extent or exhibit a multi-layered
structure.  To account for this, we have substituted approximate
but more realistic 
forms for the potential
in the diagrams of Fig.~\ref{figure1} more appropriate to
2D electrons in metal-oxide-semiconductors, in thin metallic films, or 
in layered superlattices.  Similar results are found for systems where realistic thicknesses and
superlattice spacings are used except that
the overall enhancement from these diagrams compared to \piz 
is reduced depending on the thickness (or spacing between the planes), 
dielectric constant, and the effective mass of the system. 
Our results extend to other two-dimensional 
systems and apply to the 2D electron or hole gas 
systems found in metal-oxide-semiconductors. The debate over
the nature of the metallic state in the recently observed metal-insulator 
transition of 2D electron or hole semiconductor systems still ensues with 
explanations ranging from superconductivity to more exotic 
non-Fermi liquid behavior.  The present results may bear on 
(and lend support to) explanations of the former type.  More importantly, 
we note that accurate data at $2 k_F$ is of considerable interest,
particularly from Monte Carlo simulations.  Large error bars generally 
accompany polarization data from Monte Carlo simulations in the intermediate 
regime around $2 k_F$ and make difficult an accurate determination 
of quantities such as local field factor corrections and corresponding 
effective interactions that could lay to rest debates concerning 
the likelihood of superconductivity or other electronic instabilities 
in these systems. 

\bigskip
This work
is supported by the National Science Foundation under Grant
DMR-9988576.

%The general three loop diagram \lam and the effective interaction vertex \gam.

\ecols
%\begin{figure}
%\centerline{\psfig{figure=diffuson.eps,width=8.5cm}}
%\refstepcounter{figure}
%\label{DiffusonFig}
%\smallskip
%{\small FIG. \ref{DiffusonFig}. (a) direct contribution to the
%Andreev conductance. The trace appearing in Eq. (\ref{TEFormula})
%is represented by the bubble diagram.  The double lines indicate
%the averaged Green function $\langle \GG \rangle$.  The index $n$
%corresponds to a mode coupled to the normal contact and should be
%summed over. (b) Diffusion contribution to the Andreev
%conductance. (c) Dyson equation for the $4M \times 4M$ matrix D.
%Note that upper and lower branches need to have matching mode
%indices, but not matching particle-hole indices. }
%\end{figure}
%begin{mathletters}
%labels 1.1(a), 1.1(b),...
%\end{mathletters}

\end{document}